\documentclass[final,3p,times]{elsarticle}
\usepackage{amssymb}
\usepackage{graphicx,setspace}
\usepackage{psfrag}
\usepackage{amsfonts}
\expandafter\let\csname equation*\endcsname\relax
\expandafter\let\csname endequation*\endcsname\relax
\usepackage{amsmath}
\usepackage{amssymb, amsthm}
\usepackage{mathrsfs}
\usepackage{epstopdf}
\usepackage{float}
\usepackage{lineno,hyperref}
\usepackage{color}
\usepackage{subfigure}
\usepackage{amsmath}
\usepackage{dsfont} 
\usepackage{amssymb} 
\usepackage{graphicx,color}
\usepackage{extarrows}

\usepackage{graphicx}

\usepackage{epstopdf}

\journal{arXiv}
\begin{document}
\newtheorem{definition}{Definition}[section]
\newtheorem{lemma}{Lemma}[section]
\newtheorem{remark}{Remark}[section]
\newtheorem{theorem}{Theorem}[section]
\newtheorem{proposition}{Proposition}
\newtheorem{assumption}{Assumption}
\newtheorem{example}{Example}
\newtheorem{corollary}{Corollary}[section]
\def\ep{\varepsilon}
\def\Rn{\mathbb{R}^{n}}
\def\Rm{\mathbb{R}^{m}}
\def\E{\mathbb{E}}
\def\hte{\hat\theta}
\renewcommand{\theequation}{\thesection.\arabic{equation}}
\begin{frontmatter}

\title{L\'evy noise drives an exponential acceleration in transition rates within metastable systems}

\author{Shenglan Yuan\corref{cor1}\fnref{addr1}}\ead{shenglanyuan@gbu.edu.cn}\cortext[cor1]{Corresponding author}

\address[addr1]{\rm Department of Mathematics, School of Sciences, Great Bay University, Dongguan 523000, China }

\begin{abstract}
L\'evy noise influences diverse non-equilibrium systems across scales, including quantum devices, active biological matter, and financial markets. While such noise is pervasive, its overall impact on activated transitions between metastable states remains unclear, despite prior studies of specific noise forms and scaling limits. In this work, we introduce a unified framework for L\'evy noise defined by its finite intensity and independent stationary increments. By identifying the most probable transition paths as minimizers of a stochastic action functional, we derive analytical scaling laws for escape rates under weak noise, thereby extending the classical Arrhenius law. Our results demonstrate that L\'evy noise universally enhances escape efficiency by reducing the effective potential barrier compared to Gaussian noise with equivalent intensity. Strikingly, even vanishingly weak L\'evy noise can exponentially increase escape rates across a broad range of amplitude distributions. This phenomenon arises from discontinuous most probable transition paths, where escape occurs via finite jumps. We validate these paths through the cumulant-generating function, a path integral representation, the mean first passage time and numerical simulations. Our findings reveal fundamental distinctions in escape dynamics under thermal and athermal fluctuations, suggesting new strategies to optimize switching processes in metastable systems through engineering noise properties.
\end{abstract}


\end{frontmatter}

\section{Introduction}

Many important processes in science involve the escape of a particle over an energy barrier \cite{H}. For decades, the theoretical framework for understanding the dynamics of metastable systems--from chemical reactions and protein folding to magnetic bit switching--has been dominated by  random fluctuations modeled as Gaussian (white) noise.
Kramers \cite{K} established the foundational theory of escape from metastable states under Gaussian noise, yielding the famous Arrhenius-Kramers law.

As scientific inquiry progressed into more complex systems (such as the cytoplasmic environment of living cells, turbulent fluid flows, and financial markets), it became increasingly clear that Gaussian noise provides an incomplete description. Empirical data revealed fluctuations characterized by heavy-tailed statistics and intermittency \cite{CWY,KRS,YSD}. The probability of observing large, abrupt changes (``jumps") is vastly higher than predicted by a Gaussian distribution \cite{CMGKT,VVP,YBD}. Periods of small quiescent fluctuations are punctuated by sudden large-scale events, meaning that extreme events are not statistical outliers but an integral part of the dynamics. Ankerhold \cite{A} obtained specific results for the rate asymmetry due to the third moment of current noise, enabling the analysis of experimental data and the optimization of  detection circuits, thereby applying non-Gaussian noise theory to a concrete physical system  and demonstrating its relevance for quantum device physics.
Baule and Sollich \cite{BS} identified a new universality class of non-Gaussian noises for which escape paths are dominated by large jumps.
Sung, \emph{et al}. \cite{YS} highlighted the modern relevance of characterizing non-Gaussian noise in quantum technologies.  Yuan and Bl\"{o}mker \cite{YB} established approximations via modulation or amplitude equations
for nonlinear stochastic partial differential equation driven by cylindrical $\alpha$-stable L\'evy processes.

A pivotal theoretical insight, emerging strongly in the early 2000s, was that incorporating L\'evy noise into models of metastable systems fundamentally alters the escape mechanism \cite{KCK,YZD}.  In this framework, the system is no longer constrained to a slow, diffusive climb. Instead, the heavy tails of L\'evy noise provide a  novel, fast ``leaping" pathway for escape \cite{P,SY}.  A single large noise pulse can directly propel the particle from the bottom of one potential well over the barrier into another well, rendering the transition discontinuous rather than continuous. Dubkov, Guarcello, Spagnolo \cite{DGS} characterized barrier-crossing events for superdiffusion using symmetric L\'evy flights. Their analytical results revealed an enhancement of the mean residence time in the metastable state due to L\'evy noise. Dybiec, Gudowska-Nowak, and H\"{a}nggi \cite{DGH} explored escape phenomena induced by L\'evy noise from a metastable state, comparing it directly with Kramers' theory and discussing the role of different stochastic calculi (e.g., Marcus vs. It\^{o}). Imkeller and Pavlyukevich \cite{IP} considered a dynamical system  driven by low-intensity L\'evy noise and showed that the perturbed system exhibits metastable behaviour, i.e.,it resembles a Markov jump process taking values in the local minima of the potential on a proper time scale.

For high barriers, the probability of a diffusive climb becomes astronomically small. However, the probability of a large jump, while still rare, decays only algebraically (as a power law). Consequently, for a wide range of parameters, the leaping pathway dominates, and the overall escape time becomes orders of magnitude shorter than the Kramers prediction.
Chechkin, Sliusarenko, Metzler, and Klafter \cite{CSMK} clearly demonstrated and calculated the bifurcation in the escape mechanism for L\'evy flights in a bistable potential, showing the transition from slow barrier-dominated escape to  fast noise-dominated leaping as the L\'evy index $\alpha$ decreases. Fogedby \cite{F} studied L\'evy flights characterized by a step index in a quenched isotropic short-range random force field to one-loop order.

This paradigm shift moves beyond the restrictive Gaussian assumption towards a more general and realistic theory of non-equilibrium statistical mechanics. It forces a re-evaluation of core concepts like transition states, reaction coordinates, and escape paths, supplementing the ``path of least action" with the ``path of the least improbable large jump". It also necessitates new mathematical tools, as standard methods like the Fokker-Planck equation must be replaced by fractional Fokker-Planck equations to handle the non-local nature of L\'evy jumps. Huang, \emph{et al}. \cite{YH} developed a path integral  method to obtain the most probable transition path for stochastic dynamical systems with symmetric $\alpha$-stable L\'evy motion or Brownian motion. Touchette \cite{T} presented large deviation theory, which forms the mathematical backbone for deriving weak-noise escape rates in the path-integral framework. Abebe, \emph{et al}. \cite{ATA} investigated the most probable phase portrait of a stochastic single-species
model with Allee effect driven by both non-Gaussian and Gaussian noise  using the non-local Fokker-Planck equation.

Metastable systems under L\'evy noise have applications across multiple disciplines \cite{AT,BLMV,VRDL,YW,ZYS}, including biophysics, nanotechnology, materials science, climate science, and economics. Some proteins fold or change shape faster than  traditional models predict. L\'evy noise, potentially arising from crowded active cellular environment, could provide a physical mechanism for this accelerated search through the energy landscape. Reactions in viscous solvents, membranes, or porous materials may experience non-Gaussian fluctuations, leading to significant deviations from Arrhenius law. The motion of molecular motors along filaments, exhibiting sudden jumps and pauses, may be better described by L\'evy-driven models. Ariga, Tateishi, Tomishige, and Mizuno \cite{ATTM} observed the movement of single kinesin molecules  under noisy external forces mimicking intracellular active fluctuations and found that kinesin accelerates under noise, especially under large hindering load. Qiao and Yuan  \cite{QY} considered a non-autonomous predator-prey model with diseased prey subject to
L\'evy noise and examined the asymptotic properties of the solution. At the nanoscale, components are more susceptible to large fluctuations. Understanding L\'evy-noise-induced transitions is crucial for predicting failure rates (e.g., in transistors or memory bits) and designing more robust devices. Guarcello, Valenti, Carollo, and Spagnolo \cite{GVCS} demonstrated the impact of L\'evy noise in nonlinear systems and Josephson junctions.
Climate system transitions between stable states (e.g., ice age to interglacial period) can be viewed as escapes from metastable states. The heavy-tailed jump-like forcing from volcanic eruptions or rapid greenhouse gas release can be modeled as L\'evy noise, potentially triggering exponentially faster transitions than smooth Gaussian models predict. Yuan, Li, and Zeng \cite{YLZ} characterized stochastic bifurcations and tipping phenomena of insect
outbreak dynamical systems driven by $\alpha$-stable L\'evy processes. Similarly, metastable models of market stability can exhibit ``Minsky moments" where a single large shock (a L\'evy jump) can cause a rapid crash, bypassing the slow diffusive warning signs. Zulfiqar, \emph{et al}. \cite{HY} constructed slow manifolds with exponential
tracking properties for nonlocal fast-slow stochastic evolutionary systems with stable L\'evy
noise and presented examples with numerical simulations.

This research aims to solve fundamental problems at the intersection of stochastic dynamics and non-equilibrium systems  by developing a unified theoretical framework for analyzing activated transitions (escapes from metastable states) in systems driven by  general  L\'evy noise, defined by finite intensity and independent stationary increments.
The core problem is to determine how L\'evy noise alters the scaling laws for escape rates from metastable states in the weak-noise limit. Specifically, it investigates whether and how the classical Arrhenius law (exponential scaling with barrier height for Gaussian noise) is modified by non-Gaussian heavy-tailed fluctuations. This work seeks to uncover the physical mechanism behind any acceleration in transition rates, addressing the key question of whether escape occurs via continuous paths (like Gaussian diffusion) or through fundamentally different mechanisms such as discontinuous jumps. The research also provides quantitative tools for predicting transition rates and mean first passage times in systems subject to combined Gaussian and non-Gaussian L\'evy noise, thereby extending Kramers' theory.

This paper is organized as follows. In Section \ref{MS}, we present the metastable model and examine the dynamics of a particle moving in a bistable potential under the influence of L\'evy noise, utilizing its cumulant-generating function. In Section \ref{PI}, we develop
 a comprehensive framework to understand how non-Gaussian noise characteristics fundamentally alter escape dynamics in metastable systems, establishing precise mathematical conditions for when these effects become significant.  Key results include  the derivation of characteristic functions, a path integral formulation, a weak-noise scaling analysis, and  an escape rate analysis. In Section \ref{MFET}, we provide a comprehensive and well-structured  investigation of metastable escape dynamics under combined L\'evy noise, bridging theoretical developments of  mean first escape time  with practical numerical simulation across multiple scaling regimes. We present a parameter dependence analysis
 by incorporating both Gaussian and L\'evy jump noise components. Our findings show that synergistic noise parameters can work together to reduce the mean first escape time more than expected. The compensation drift crucially modifies the potential landscape, affecting both diffusive and jump escapes, while
 jump distributions with heavier tails are shown to dominate escape dynamics.
In Section \ref{FS}, we summarize our conclusions and discuss potential directions for future research.

\section{Metastable systems driven by L\'evy noise}\label{MS}
We consider the one-dimensional physical model governed by a conservative force with potential
$V$ and a L\'evy noise term $\eta$:
\begin{equation}\label{SM}
\dot{u}(t)=-V'(u)+\eta(t),
\end{equation}
where $\eta(t)=\eta_{\text{G}}(t)+\eta_{\text{J}}(t)$.
The Gaussian component $\eta_{\text{G}}$ is standard Langevin noise, while $\eta_{\text{J}}$ is a compensated compound Poisson process that accounts for all jumps:
\begin{equation}\label{kicks}
\eta_{\text{J}}(t)=C(t)-\lambda_0 t\mathbb{E}H.
\end{equation}
Here, $C(t)$ is a compound Poisson process:
\begin{equation}\label{compound}
C(t)=\sum_{k=1}^{N_{t}}H_{k},\quad H_{k}\sim \mu\, \text{iid and independent of}\,(N_{t})_{t\geq0},
\end{equation}
where the random jump heights $H_k$  are identically
and independently distributed with mean value $\mathbb{E}H$.
The amplitudes $H_k$ are independently drawn from a distribution $\mu$ with density
$\rho(H/h_0)/h_0$, where $h_0$ sets the height scale, and $\rho(x)$ is normalized such that $\int x^2\rho(x)dx=1$. In \eqref{compound},  the integer-valued number  of counting jumps is a Poisson
process
\begin{equation*}
N_{t}=\sum_{k=1}^{\infty}\mathds{1}_{[0,t]}(\tau_{k}),\quad\tau_{k}=\sigma_{1}+\cdots+\sigma_{k},\quad\sigma_{k}\sim \textup{Exp}(\lambda_0)\quad \textup{iid},
\end{equation*}
which increases by 1 after each
independent exponential waiting time with mean $\lambda_0$. The times $\tau_k$  are a sequence of stopping times at which jumps occur. To ensure the noise has zero mean, we subtract the resulting expectation $\mathbb{E}C(t)=\lambda_0 t\mathbb{E}H$ of the compound Poisson process
so that $\mathbb{E}\eta_{\text{J}}(t)=0$ for any distribution of jump sizes in Eq. \eqref{kicks}.

\begin{figure}[H]
 \begin{center}
 \begin{minipage}{3.5in}
 \includegraphics[width=3.5in]{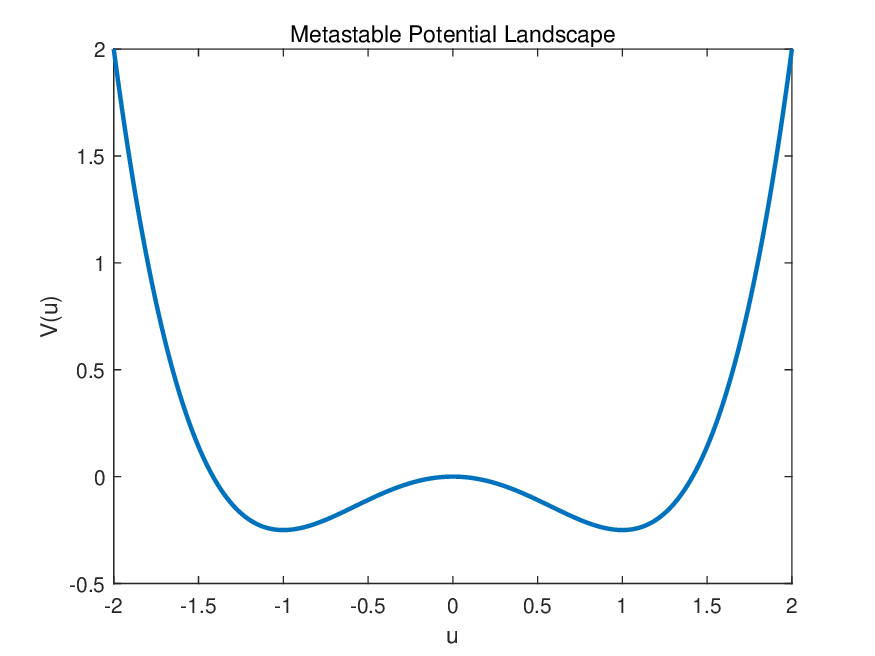}
 \end{minipage}
\caption{The bistable potential $V(u)=\frac{u^{4}}{4}-\frac{u^{2}}{2}$ possesses two symmetric metastable states at $u=\pm1$ and an energy barrier at $u=0$.}\label{MPLV}
\end{center}
\end{figure}

Metastability arises in \eqref{SM} when $V(u)$ contains two or more sufficiently deep minima, trapping the particle near one minimum for long periods, with rare noise-induced escapes to neighboring wells. Fig. \ref{MPLV}
characterizes a potential function $V(u)=\frac{u^{4}}{4}-\frac{u^{2}}{2}$, which has two symmetric metastable states at $u=\pm1$ and an energy barrier at $u=0$.

The noise properties are fully characterized by the cumulant-generating function defined as the natural logarithm of the
characteristic function:
\begin{equation}\label{cumulant}
\ln\mathbb{E}\left[\exp\left(i\int_{0}^{T}\xi(t)\eta(t)dt\right)\right]=\int_{0}^{T}\Big[\frac{D_0}{2}(i\xi)^{2}+\lambda_0\phi(ih_0\xi)\Big]dt,
\end{equation}
where
\begin{equation}\label{moment}
\phi(y)=\int(e^{xy}-xy-1)\rho(x)dx
\end{equation}
generates the moments of the non-Gaussian component. An advantage is that  the cumulant-generating function in \eqref{cumulant}
is well-defined since $\mathbb{E}\left[\exp\left(i\int_{0}^{T}\xi(t)\eta(t)dt\right)\right]$ is well-defined for all
functions $\xi(t)$ such that the integral $\int_{0}^{T}\xi(t)\eta(t)dt$ converges for almost all realizations of $\eta$.
The Gaussian term $\frac{D_0}{2}(i\xi)^2$ has variance $D_0$, while  $\lambda_0\phi(ih_0\xi)$ captures compensated compound Poisson noise. The total variance is
$\mathbb{E}(\eta(t)\eta(t'))=
(D_0+\lambda_0h^2_0)\delta(t-t')$. We restrict our analysis to symmetric noise  with $\rho(x)=\rho(-x)$. This framework  generalizes to L\'evy noise, including the $\alpha$-stable L\'evy noise cases where $\rho(x)$ diverges as $|x|^{-\alpha-1}$ for  $0<\alpha<2$. As shown by Samorodnitsky and Taqq \cite{ST}, stable distributions constitute a central subclass of L\'evy processes.

\section{Non-Gaussian escape rates via path integral}\label{PI}
For the compound Poisson noise defined in \eqref{compound}, we consider the increment
 $\bar{C}(s):= \int_{s}^{s+\Delta t}C(r)dr$ over a small time interval $\Delta t$. These increments are independent and, to first order in $\Delta t$, $\bar{C}(s)$
 takes the value $H$ with probability $\lambda_0\Delta t$ or the value
0 with probability
$1-\lambda_0\Delta t$ otherwise. The characteristic function of a single increment is therefore given by
\begin{equation*}
\mathbb{E}e^{i\xi(s)\bar{C}(s)}=\lambda_0\Delta t\mathbb{E}e^{i\xi(s)H}+(1-\lambda_0\Delta t)=1+\lambda_0\Delta t\mathbb{E}\big(e^{i\xi(s)H}-1\big).
\end{equation*}
For small
$\Delta t$, this approximates to
\begin{equation*}
\mathbb{E}e^{i\xi(s)\bar{C}(s)}\approx\exp\Big\{\lambda_0\Delta t\mathbb{E}\big(e^{i\xi(s)H}-1\big)\Big\}.
\end{equation*}

For the compensated compound Poisson noise defined in \eqref{kicks}, the constant mean $\mathbb{E}C(t)$
is subtracted. This compensation introduces an additional linear term $-i\xi(s)H$ within the exponential average in the characteristic function. Considering the increment over a small time interval $\Delta t$,
 $$\bar{\eta}_{\text{J}}(s):= \int_{s}^{s+\Delta t}\eta_{\text{J}}(r)dr=\int_{s}^{s+\Delta t}\big(C(r)-\mathbb{E}C(r)\big)dr=\bar{C}(s)-\mathbb{E}\bar{C}(s),$$
the characteristic function of this increment is then
\begin{align*}
\mathbb{E}e^{i\xi(s)\bar{\eta}_{\text{J}}(s)}&=\mathbb{E}e^{i\xi(s)\big(\bar{C}(s)-\mathbb{E}\bar{C}(s)\big)}=\frac{\mathbb{E}e^{i\xi(s)\bar{C}(s)}}{\mathbb{E}e^{i\xi(s)\mathbb{E}\bar{C}(s)}}\\
&\approx\frac{\exp\Big\{\lambda_0\Delta t\mathbb{E}\big(e^{i\xi(s)H}-1\big)\Big\}}{\exp\Big\{i\xi(s)\lambda_0\Delta t\mathbb{E}H\Big\}}\\
&=\exp\Big\{\lambda_0\Delta t\mathbb{E}\big(e^{i\xi(s)H}-i\xi(s)H-1\big)\Big\},
\end{align*}
where we have used the previous approximation for $\mathbb{E}e^{i\xi(s)\bar{C}(s)}$ and the fact that $\mathbb{E}\bar{C}(s)=\lambda_0 \Delta t \mathbb{E}H $.

Due to the independence of the Gaussian and jump components, the characteristic function of the combined noise $\bar{\eta}(s) = \bar{\eta}_{G}(s) + \bar{\eta}_{J}(s)$--which incorporates a Gaussian component with variance $D_0$--is given by
\begin{align}\nonumber
\mathbb{E}e^{i\xi(s)\bar{\eta}(s)}
&=\mathbb{E}e^{i\xi(s)(\bar{\eta}_{G}(s)+\bar{\eta}_{J}(s))}=\mathbb{E}e^{i\xi(s)\bar{\eta}_{G}(s)}\mathbb{E}e^{i\xi(s)\bar{\eta}_{J}(s)}\\ \nonumber
&=\exp\Big(-\frac{D_0}{2}\xi(s)^{2}\Delta t\Big)\exp\Big(\lambda_0\phi_0(i\xi(s))\Delta t\Big)\\\label{characteristicfct}
&=\exp\Big(-\frac{D_0}{2}\xi(s)^{2}\Delta t+\lambda_0\phi_0(i\xi(s))\Delta t\Big),
\end{align}
where $\phi_0(y)$ is defined as
\begin{equation}\label{phi0}
\phi_0(y)=\int(e^{Hy}-Hy-1)\rho_0(H)dH.
\end{equation}

For generality, the amplitude distribution $\rho_0(H)$
is expressed in terms of a scaled base distribution $\rho$:
\begin{equation*}
\rho_0(H)=\frac{1}{h_0}\rho(\frac{H}{h_0}),
\end{equation*}
where $h_0$ is a characteristic scale, and
$\rho(x)$ is normalized such that $\int x^{2}\rho(x)dx=1$.
This scaling implies $\phi_0(y)=
\phi(h_0y)$, where
\begin{equation}\label{phiu}
\phi(y)=\int(e^{Hy}-Hy-1)\rho(H)dH.
\end{equation}
Under this normalization, the L\'evy noise variance becomes $\lambda_{0}\mathbb{E}H^{2}=\lambda_{0}h_0^{2}$. Taking the continuum limit $\Delta t\rightarrow0$,  the natural logarithm of \eqref{characteristicfct}
recovers the noise cumulant generator in \eqref{cumulant}.

To construct a path integral formulation for the dynamics in \eqref{SM}, we discretize the equation using the It\^{o} convention:
\begin{equation}\label{discretization}
u(s+\Delta t)=u(s)-V'(u(s))\Delta t+\bar{\eta}(s).
\end{equation}
The probability of a trajectory
$[u]:=(u(0),u(\Delta t),\cdots,u(t))$ with fixed initial condition $u(0)$ is expressed as a product of delta functions enforcing the dynamics at each step:
\begin{equation}\label{product}
\begin{array}{l}
P[u]=\mathbb{E}\left(\prod_{s=0}^{t-\Delta t}\delta\big(u(s+\Delta t)-u(s)+V'(u(s))\Delta t-\bar{\eta}(s)\big)\right)\\
\quad\quad\,=\prod_{s=0}^{t-\Delta t}\mathbb{E}\delta\big(u(s+\Delta t)-u(s)+V'(u(s))\Delta t-\bar{\eta}(s)\big),
\end{array}
\end{equation}
where the average is over the noise increments $\bar{\eta}(s):=\int_{s}^{s+\Delta t}\eta(r)dr$.

Applying the Fourier transform to each delta function in \eqref{product} introduces auxiliary variables $\xi(s)$, leading to
\begin{equation*}
\frac{1}{2\pi}\int e^{-i\xi(s)[u(s+\Delta t)-u(s)+V'(u(s))\Delta t]}\mathbb{E}e^{i\xi(s)\bar{\eta}(s)}d\xi(s).
\end{equation*}
Substituting the characteristic function from Eq. \eqref{characteristicfct} then yields
\begin{equation}\label{Fourier}
\frac{1}{2\pi}\int e^{-i\xi(s)[u(s+\Delta t)-u(s)+V'(u(s))\Delta t]-\frac{D_0}{2}\xi(s)^{2}\Delta t+\lambda_0\phi(ih_0\xi(s))\Delta t}d\xi(s).
\end{equation}

Combining contributions from all time steps and taking the continuum limit $\Delta t\rightarrow0$, the path probability is expressed via the Martin-Siggia-Rose (MSR) action functional \cite{YD}:
\begin{equation*}
P[u]=\int e^{-S[u,\xi]}\mathcal{D}\Big[\frac{\xi}{2\pi}\Big],
\end{equation*}
where
\begin{equation*}
S[u,\xi]=\int_{0}^{t}\Big\{i\xi(s)\big(\dot{u}(s)+V'(u(s))\big)+\frac{D_0}{2}\xi(s)^{2}-\lambda_0\phi(ih_0\xi(s))\Big\}ds.
\end{equation*}
This formulation encapsulates both Gaussian and non-Gaussian noise contributions within a unified path integral framework.

The foundational Kramers escape rate for systems driven by Gaussian noise
$(\lambda_0=0)$ is derived from
 large deviation theory in the weak-noise limit $D_0\rightarrow0$. To formalize this weak-noise behavior, a dimensionless parameter $\epsilon$
is introduced, rescaling the noise strength as
\begin{equation}\label{rescaleD}
D_0=D\epsilon,
\end{equation}
where $\epsilon\rightarrow0$ defines the weak-noise limit. For $\lambda_0=0$, the stationarity condition $\delta S/\delta \xi(s)=0$ yields $D\epsilon \xi=-i(\dot{u}+V')$. This relation necessitates the scaling
$\xi=\tilde{\xi}/\epsilon$ to maintain  a consistent equation $D\tilde{\xi}=-i(\dot{u}+V')$. In the absence of the non-Gaussian term (i.e., $\lambda_0=0$), this $\epsilon^{-1}$ scaling ensures that deviations from the most probable transition path are exponentially suppressed for small $\epsilon$, thereby recovering the classical result as $\epsilon$ approaches zero.

For non-zero $\lambda_0$, the rescaled action becomes
\begin{equation}\label{newaction}
S[u,\tilde{\xi}]=\frac{1}{\epsilon}\int_{0}^{t}\Big\{i\tilde{\xi}[\dot{u}+V'(u)]+\frac{D\tilde{\xi}^{2}}{2}-\lambda_0\epsilon\phi(ih_0\frac{\tilde{\xi}}{\epsilon})\Big\}ds.
\end{equation}
The challenge lies in identifying a scaling that preserves non-Gaussian noise contributions in the weak-noise limit. The form of the term
 $\lambda_0\epsilon\phi(ih_0\tilde{\xi}/\epsilon)$ in \eqref{newaction} suggests the scaling $\lambda_0=\lambda/\epsilon$ and $h_0=\epsilon h$.

To generalize, we consider scaling exponents
\begin{equation}\label{scalingexponents}
\lambda_0=\lambda/\epsilon^{\mu},\quad h_0=\epsilon^{\nu}h,
\end{equation}
and expand $\phi$ in Eq. \eqref{newaction}:
\begin{equation}\label{Expanding}
\phi(ih_0\tilde{\xi}/\epsilon)=\frac{(i\tilde{\xi})^{2}}{2!}\frac{h_0^{2}\mathbb{E}H^{2}}{\epsilon^{2}}+\frac{(i\tilde{\xi})^{3}}{3!}\frac{h_0^{3}\mathbb{E}H^{3}}{\epsilon^{3}}+\cdots.
\end{equation}
Each term of order $O(\tilde{\xi}^n)$ in $\lambda_0\epsilon\phi$
then scales as
$\epsilon^{1-\mu+n(\nu-1)}$.
Three distinct regimes emerge as $\epsilon\rightarrow0$:
\begin{description}
  \item[\textbf{Regime I}] $(\mu>2\nu-1,\nu<1)$: All terms ($n\geq2$) in $\tilde{\xi}$ diverge as $\epsilon\rightarrow0$.
  \item[\textbf{Regime II}] $(\mu<2\nu-1,\nu\leq1 \,\text{or}\,\, \mu>2\nu-1,\nu\geq1)$: Some higher-order terms diverge as $\epsilon\rightarrow0$.
  \item[\textbf{Regime III}] $(\mu<2\nu-1,\nu>1)$: All terms vanish, reducing to Gaussian noise $(\lambda_0=0)$.
\end{description}
Only along the critical line $\mu=2\nu-1$ with $\nu>1$ does the $\tilde{\xi}^2$ term survive, but this reduces to effective Gaussian noise since
$\lambda_0h_0^2\propto\epsilon\rightarrow0$.

The unique preservation of non-Gaussianity occurs for
$\mu=\nu=1$, where all terms in $\phi$ remain finite. This scaling ensures that the noise variance
$D_0+\lambda_0h_{0}^{2}=(D+\lambda h^{2})\epsilon\propto\epsilon$
while retaining all cumulants. The action simplifies to
\begin{equation}\label{simpleaction}
\tilde{S}[u,\tilde{\xi}]=\int_{0}^{t}\Big\{i\tilde{\xi}[\dot{u}+V'(u)]+\frac{D\tilde{\xi}^2}{2}-\lambda\phi
(ih\tilde{\xi})\Big\}ds,
\end{equation}
with path probabilities given by
\begin{equation}\label{probabilities}
P[\xi]=\int e^{-\tilde{S}[u,\tilde{\xi}]/\epsilon}\mathcal{D}\Big[\frac{\tilde{\xi}}{2\pi\epsilon}\Big],\quad \text{where}\,\, \tilde{\xi}=\epsilon\xi\, \text{and}\, \tilde{S}[u,\tilde{\xi}]=\epsilon S[u,\tilde{\xi}].
\end{equation}
The scaling $\lambda_0=\lambda/\epsilon$, $h_0=\epsilon h$ defines a consistent weak-noise limit that retains the full hierarchy of non-Gaussian noise effects, generalizing Kramers' theory beyond Gaussian assumptions.

Given the large deviation formulation of the path probability, the propagator for the dynamics described by Eq. \eqref{SM}, representing the probability of transitioning from an initial state
$u(0)$ to a final state $u(t)$, can be expressed as a path integral over all trajectories connecting these endpoints. In the weak-noise limit ($\epsilon\rightarrow0$), this propagator is predominantly governed by the path that renders the action in Eq. \eqref{simpleaction} stationary, typically found by solving the Euler-Lagrange equations for $u(s)$ and $\tilde{\xi}(s)$.
However, these equations assume continuous paths, which may not exist for certain types of non-Gaussian noise.
To address this, we first eliminate $\tilde{\xi}$ from Eqs. \eqref{simpleaction}-\eqref{probabilities} via saddle-point integration in the $\epsilon\rightarrow0$ limit. This procedure involves discretizing time into intervals $\Delta t$, taking the limit $\epsilon\rightarrow0$ first, followed by $\Delta t\rightarrow0$.

The
stationarity condition
\begin{equation}\label{stationarity}
0=i[\dot{u}+V'(u)]+D\tilde{\xi}-i\lambda h\phi'(ih\tilde{\xi})
\end{equation}
reveals that
 $\tilde{\xi}$ is purely imaginary at the saddle point. Substituting $q=i\tilde{\xi}$ into the integrand $i\tilde{\xi}[\dot{u}+V'(u)]+\frac{D\tilde{\xi}^2}{2}-\lambda\phi
(ih\tilde{\xi})$ from Eq. \eqref{simpleaction},
 we obtain the Lagrangian density, which characterizes the action's contribution:
\begin{align}\nonumber
\mathcal{L}(f)&=\max_{q}\{qf-\frac{D}{2}q^2-\lambda\phi(hq)\} \\ \label{LF}
              &=\max_{q}\{qf-\psi(q)\},
\end{align}
where $f:=\dot{u}+V'(u)$ and $\psi(q):= \frac{D}{2}q^2 + \lambda\phi(hq)$.

Two key properties of $\phi$, $\psi$, and $\mathcal{L}$ are their convexity and symmetry. First, $\phi$ is convex, and consequently, $\mathcal{L}$ is also convex as a Legendre transform. Second, for a symmetric $\rho$, the functions $\phi$, $\psi$, and $\mathcal{L}$ are symmetric with minima at zero; this results in vanishing odd moments for the noise and even moments bounded by $\phi(y)\geq y^2/2$, which in turn leads to $\psi(q)\geq(D+\lambda h^2)q^2/2$.

Now we analyze the exit path from a metastable state $u_{\text{min}}$,
 situated at the minimum of a potential $V$, over the nearest barrier to arrive at $u_{\text{max}}>u_{\text{min}}$.

For systems driven by Gaussian noise, the escape rate $r$ scales asymptotically as
$r\cong Ce^{-S_{\text{min}}/D}$ in the limit of small noise intensity $D$, where $S_{\text{min}}$
represents the minimal action governing the transition. This result parallels semiclassical calculations of quantum tunneling.

For non-Gaussian noise, the escape dynamics are governed by large deviation theory in the limit $\epsilon\rightarrow0$. The effective energy barrier is determined by the minimum action
\begin{equation}\label{minimumaction}
S_{\text{min}}=\lim_{T\rightarrow\infty}\min\limits_{[u]}\int_{0}^{T}\mathcal{L}(\dot{u}+V'(u))dt.
\end{equation}
where the minimization is performed over all paths $u(t)$ connecting $u(0)=u_{\text{min}}$ to $u(T)=u_{\text{max}}$.

To make progress in determining $S_{\text{min}}$, we can employ a geometric reformulation in the $(u,v)$-plane, where $v=\dot{u}$.
By parametrizing paths in the $(u,v)$-plane, the action becomes
$\int\mathcal{L}(v+V'(u))/|v|du$. For each position
$u$, the velocity $v$ is determined by minimizing the quantity $\mathcal{L}(v+V'(u))/|v|$.

The
key minimization condition is that if the minimum occurs at a finite velocity
$v$, it satisfies
\begin{equation*}
\mathcal{L}(f)=v\mathcal{L}'(f),\quad \text{where}\,\, f=v+V'(u).
\end{equation*}
Using the Legendre transform relation $\mathcal{L}(f)=q^{*}f-\psi(q^{*})$, and consequently $\mathcal{L}'(f)=q^{*}$
(with $q^{*}=\text{argmax}_q[qf-\psi(q)]$), the minimum action simplifies to
\begin{equation}\label{action}
S_{\text{min}}=\int_{u_{\text{min}}}^{u_{\text{max}}}q^{*}(V'(u))du,
\end{equation}
where the function $q^{*}(V')$ is defined implicitly as the solution to
\begin{equation}\label{kstar}
V'(u)=\psi(q^{*})/q^{*}.
\end{equation}
Here, $\psi$ is the cumulant-generating function of the noise.

Based on the inverse Legendre transform relation $\psi'(q^{*})=f$, the velocity along the most probable transition path is given by
\begin{equation}\label{inverseLegendre}
v=\dot{u}=\psi'(q^{*})-V'(u).
\end{equation}
This defines a velocity function
$\dot{u}=F(V'(u))$ that characterizes the shape of the escape path, where $F$ is determined by combining this result with Eq. \eqref{kstar}.

Drawing a comparison to classical mechanics, where $\partial S/\partial u=p$,
the term $q^{*}$ in Eq. \eqref{action} acts analogously to a momentum. The Hamiltonian
$\mathcal{H}=q^{*}\dot{u}-\mathcal{L}=-q^{*}V'(u)+\psi(q^{*})$
vanishes ($\mathcal{H}=0$) for minimum-action paths of infinite duration in the limit $T\rightarrow\infty$. However, this Hamiltonian framework  breaks down if the most probable transition path contains discontinuous jumps, as $\dot{u}$ becomes undefined. Even in such cases, the minimization approach for $\mathcal{L}(v+V'(u))/|v|$ remains valid.

The analysis of the effective energy barrier $S_{\text{min}}$ for non-Gaussian noise reveals notable contrasts with the Gaussian regime.
Non-Gaussian noise universally induces faster escape rates, providing exponential acceleration since the escape rates scale as $\exp(-S_{\text{min}}/\epsilon)$. Specifically, this acceleration is quantified by the inequality $S_{\text{min}}< 2\Delta V/(D+\lambda h^2)$.

The reduction of the effective barrier arises from a fundamental mechanistic difference.
Non-Gaussian noise enables discontinuous jumps, which alter the escape pathway.
Mathematically, this is rooted in the inequality $\psi(q)\geq(D+\lambda h^2)q^2/2$,
which implies $q^{*}\leq
2V'/(D+\lambda h^2)$. Substituting this bound into the action integral in Eq. \eqref{action} directly yields the barrier reduction $S_{\text{min}}<2\Delta V/(D+\lambda h^2)$.

For the small-noise limit analysis ($\lambda\rightarrow0$),  the equation \eqref{kstar} governing $q^{*}$ simplifies to
\begin{equation}\label{Vq}
V'(u)=\frac{D}{2}q^{*}+\lambda\frac{\phi(hq^{*})}{q^{*}}.
\end{equation}
To explain the decrease in $S_{\text{min}}$ with
$\lambda$ and $h$, we rewrite Eq. \eqref{Vq} as
\begin{equation}\label{kV}
\tilde{q}^{\ast}=hq^{\ast},\quad V'(u)=\frac{D}{2h}\tilde{q}^{\ast}+\lambda h\frac{\phi(\tilde{q}^{\ast})}{\tilde{q}^{\ast}}.
\end{equation}
Both terms on the right are positive. If either prefactor ($D/h$ or $\lambda h$) is sufficiently large,
$\tilde{q}^{\ast}$ becomes small. For small $\tilde{q}^{\ast}$, $\phi(y)\approx y^2/2$, yielding $\tilde{q}^{\ast}=2V'/(Dh^{-1}+\lambda h)$. Substituting $q^{\ast}=\tilde{q}^{\ast}/h$, the minimum action $S_{\text{min}}$ from Eq. \eqref{action}
reduces to its Gaussian value $S_{\text{min}}\approx 2\Delta V/(D+\lambda h^2)$, and the most probable transition path adopts a Gaussian profile, as $\psi(q)\approx(D+\lambda h^2)q^2/2$ and Eq. \eqref{inverseLegendre}
lead to $\dot{u}=V'(u)$. Gaussian behavior dominates when $D/h\gg1$ or $\lambda h\gg1$. Non-Gaussian effects emerge in the regime where $D\ll h\ll1/\lambda$, which  is only possible if
$\lambda\ll1$.

\begin{figure}[H]
\begin{center}
  \begin{minipage}{2.13in}
\leftline{(a)}
\includegraphics[width=2.13in]{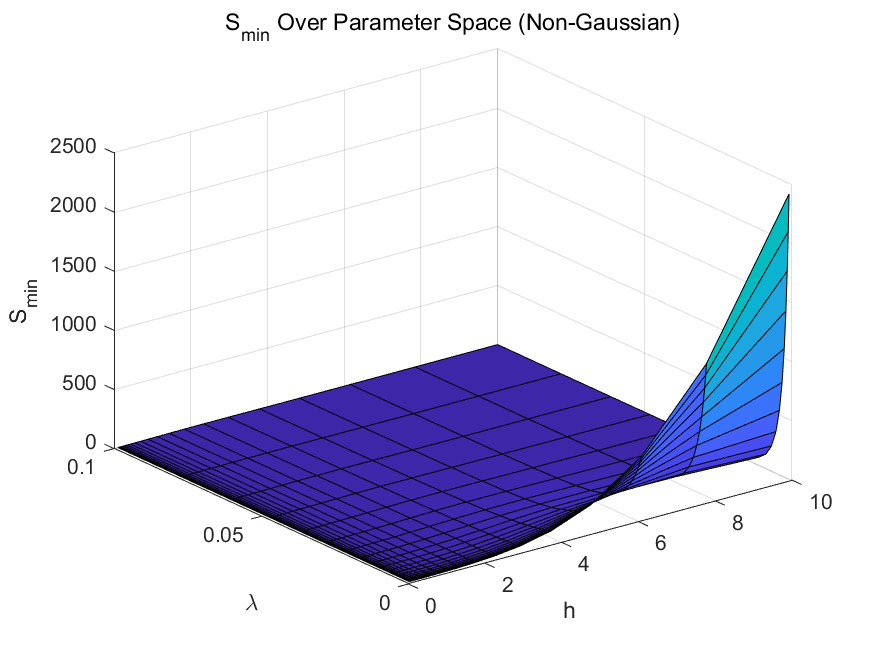}
\end{minipage}
\hfill
\begin{minipage}{2.13in}
\leftline{(b)}
\includegraphics[width=2.13in]{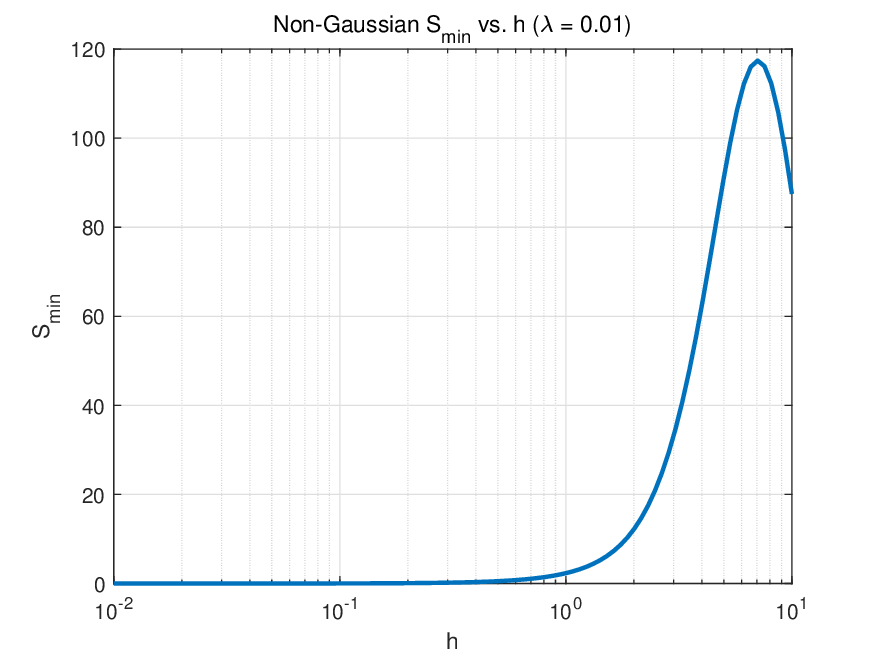}
\end{minipage}
\hfill
  \begin{minipage}{2.13in}
\leftline{(c)}
\includegraphics[width=2.13in]{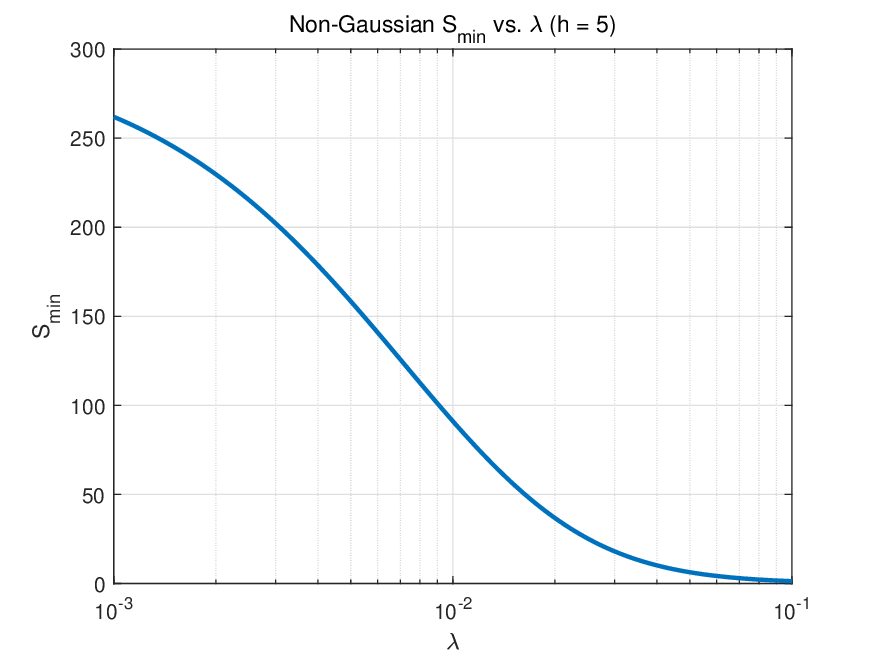}
\end{minipage}
\caption{(a) The minimum action $S_{\text{min}}$  over a parameter grid of $h$ and $\lambda$; (b) The minimum action $S_{\text{min}}$ as a function of $h$ for a fixed $\lambda=0.01$; (c) The minimum action $S_{\text{min}}$ as a function of $\lambda$  for a fixed $h=5$.}\label{Minimumaction}
\end{center}
\end{figure}

Fig. \ref{Minimumaction}(a) studies the minimum action $S_{\text{min}}$ in the parameter space of $h$ and $\lambda$ for the physical system \eqref{SM} with Gaussian and non-Gaussian contributions. Fig. \ref{Minimumaction}(b) shows the non-Gaussian action $S_{\text{min}}$ as a function of $h$ for a fixed $\lambda=0.01$. Fig. \ref{Minimumaction}(c) characterizes the physical behavior of $S_{\text{min}}$  across different $\lambda$ regimes for a fixed value of $h=5$.

\section{Mean first passage time}\label{MFET}
The mean first passage time (MFPT) for a particle to escape a metastable well governed by system \eqref{SM} with combined L\'evy noise, $\eta(t)=\eta_{\text{G}}(t)+\eta_{\text{J}}(t)$, can be derived under the assumption of a deep potential well and rare escape events. The total escape rate $\Gamma$ is the sum of independent contributions from the continuous Gaussian noise $\Gamma_{\text{G}}$, and the discontinuous jump-like L\'evy noise $\Gamma_{\text{J}}$, leading to $\text{MFPT} = 1/\Gamma$. The Gaussian component is the classic Kramers rate, $$\Gamma_{\text{G}}=\frac{\sqrt{|V''(u_{\text{min}})V''(u_{\text{max}})|}}{2\pi}\exp\left(-\frac{2\Delta V}{D_0}\right),$$ which depends on the curvature of the potential at the well bottom $u_{\text{min}}$ and barrier top $u_{\text{max}}$, the barrier height $\Delta V=V(u_{\text{max}})-V(u_{\text{min}})$, and the Gaussian noise intensity $\frac{D_0}{2}$.
In contrast, the L\'evy jump component $$\Gamma_{J}=\lambda_{0}\int_{\Delta u/h_0}^{\infty}\rho(x)dx$$ accounts for escapes triggered by large jumps. Here,  $\lambda_{0}$ is the jump arrival rate, and the integral represents the probability that a single jump of scale $h_0$ exceeds the spatial distance $\Delta u=u_{\text{max}}-u_{\text{min}}$ required to surmount the barrier directly. Thus, the full MFPT is given by the reciprocal of the sum of these two mechanisms, providing a comprehensive extension of Kramers' theory that captures both continuous diffusion over the barrier and discontinuous jumps that can induce immediate escape.

The total escape rate $\Gamma=\Gamma_{\text{G}}+\Gamma_{\text{J}}$ for a particle in a metastable well under L\'evy noise is derived by combining two mechanisms: diffusive escape over the potential barrier (driven by Gaussian noise) and direct escape triggered by large jumps (from the compound Poisson noise).

The derivation for the Gaussian component $\Gamma_{\text{G}}$ begins with the Langevin equation $\dot{u}(t)=-V'(u)+\eta_{\text{G}}(t)$, where $\mathbb{E}(\eta_{\text{G}}(t)\eta_{\text{G}}(t'))=D_0\delta(t-t')$.
Under the key assumptions of a deep well (with a high barrier $\Delta V=V(u_{\text{max}})-V(u_{\text{min}})\gg \frac{D_0}{2}$) and a quasi-stationary probability distribution $p(u)\approx\mathcal{N}e^{-2V(u)/D_0}$ near the minimum $u_{\text{min}}$, the standard Kramers' rate is recovered. This involves solving the corresponding Fokker-Planck equation
\begin{equation}
\frac{\partial p}{\partial t}=\frac{\partial}{\partial u}\Big[V'(u)p+\frac{D_0}{2}\frac{\partial p}{\partial u}\Big]
\end{equation}
with absorbing boundary conditions at the barrier top $u_{\text{max}}$ and reflecting conditions at the well bottom $u_{\text{min}}$, leading to the solution
\begin{equation*}
\Gamma_{\text{G}}=\frac{\sqrt{|V''(u_{\text{min}})V''(u_{\text{max}})|}}{2\pi}\exp\left(-\frac{2\Delta V}{D_0}\right),
\end{equation*}
where the curvatures $V''(u_{\text{min}})>0$ (well curvature) and $V''(u_{\text{max}})<0$ (barrier curvature) characterize the well and barrier.

The non-Gaussian jump component $\Gamma_{\text{J}}$ is derived from the compensated compound Poisson process
\begin{equation*}
\eta_{\text{J}}(t)=\sum_{j=1}^{N_t}H_j\delta(t-\tau_j)-\lambda_0 t\mathbb{E}[H],
\end{equation*}
where $\lambda_0$ is the jump arrival rate, and the jump sizes $H_j$ are independently and identically distributed according to a distribution $\mu$ with density $\frac{1}{h_0} \rho\left(\frac{h}{h_0}\right)$. Here, $\rho$ is a base density function normalized such that $\int x^2\rho(x)dx = 1$.

The critical assumptions are that the particle is typically located near the minimum
$u_{\text{min}}$ due to metastability when a jump occurs, and that a single jump from $u_{\text{min}}$ to $u_{\text{min}}+H_j$ is sufficient for escape if its magnitude exceeds the spatial distance to the barrier, i.e., $H_j>\Delta u=u_{\text{max}}-u_{\text{min}}$. The probability that a single jump exceeds $\Delta u$ is
\begin{equation*}
\mathbb{P}(H_j>\Delta u)=\int_{\Delta u}^{\infty}\frac{1}{h_0}\rho\left(\frac{h}{h_0}\right)dh.
\end{equation*}
Changing the integration variables  to $x=h/h_0$ yields
\begin{equation*}
\mathbb{P}(H_j>\Delta u)=\int_{\Delta u/h_0}^{\infty}\rho(x)dx.
\end{equation*}
Multiplying this probability by the jump arrival rate gives the jump-induced escape rate:
\begin{equation*}
\Gamma_{J}=\lambda_0\mathbb{P}(H_j>\Delta u)=\lambda_0\int_{\Delta u/h_0}^{\infty}\rho(x)dx.
\end{equation*}

The additivity of the total escape rate $\Gamma=\Gamma_{\text{G}}+\Gamma_{\text{J}}$
is justified by the statistical independence of the two escape mechanisms (continuous diffusion over the barrier and discontinuous jump-induced escape) and the rarity of the events (the probabilities of escape are small). The probabilities of escape from either process in a small time interval $\Delta t$ are additive: $\Gamma\Delta t=\Gamma_{\text{G}}\Delta t+\Gamma_{\text{J}}\Delta t$.

The final expression for the total escape rate is therefore
\begin{equation*}
\Gamma=\frac{\sqrt{|V''(u_{\text{min}})V''(u_{\text{max}})|}}{2\pi}\exp\left(-\frac{2\Delta V}{D_0}\right)+\lambda_{0}\int^{\infty}_{\Delta u/h_0}\rho(x)dx.
\end{equation*}

The compensation drift $-\lambda_0 t\mathbb{E}[H]$
in the jump noise is subdominant near the well bottom $u_{\text{min}}$ and can be incorporated into an effective potential $V_{\text{eff}}(u):=V(u)+(\lambda_0\mathbb{E}[H])u$ for the Gaussian component. However, this drift does not interfere with the instantaneous escape caused by a sufficiently large jump, nor do Gaussian fluctuations significantly alter jump-triggered escape.

Key physical insights reveal that $\Gamma_G$ decays exponentially with $2\Delta V/D_0$,
dominating when the barrier is high and jumps are small. In contrast, $\Gamma_{J}$ depends on the tail of the distribution $\rho(x)$ and becomes dominant for heavy-tailed jump distributions. The compensation term effectively shifts the potential to $V_{\text{eff}}(u)$, which is relevant for calculating the diffusive escape rate but does not affect the jump-induced escape, as a sufficiently large jump causes an instantaneous transition over the barrier regardless of this small deterministic drift.

The mean first passage time is given by
\begin{equation*}
\text{MFPT}=\frac{1}{\frac{\sqrt{|V_{\text{eff}}''(u_{\text{min}})V_{\text{eff}}''(u_{\text{max}})|}}{2\pi}\exp\left(-\frac{2\Delta V_{\text{eff}}}{D_0}\right)+\lambda_{0}\int^{\infty}_{\Delta u/h_0}\rho(x)dx}.
\end{equation*}
This expression extends classical Kramers' theory to include L\'evy noise, capturing both continuous diffusion and discontinuous jumps in escape dynamics.

Table \ref{Table1} presents simulation results for the escape dynamics of Eq. \eqref{SM} with the potential $V(u)=\frac{u^{4}}{4}-\frac{u^{2}}{2}$ and a Rayleigh jump distribution $\rho(x)=2xe^{-x^2}$. Using the parameters $D_0=0.05$, $\lambda_0=0.1$, and $h_0=1$, the effective potential becomes $V_{\text{eff}}(u)=V(u)+(\lambda_0EH)u$. Solving $u^3-u+0.0886=0$ yields a new well bottom at $u_{\text{min}}\approx-1.0443$. The corresponding effective barrier height is $\Delta V_{\text{eff}}=V_{\text{eff}}(0)-V_{\text{eff}}(-1.0443)\approx 0.34043$. Under these conditions,  the simulated MFPT is approximately 32.5 time units.

For the case of higher Gaussian noise ($D_0=0.1$) with parameters
$\lambda_0=0.1$ and $h_0=1$, the compensation drift significantly alters the potential landscape. It shifts the well bottom to $u_{\text{min}}\approx 1.0416$ and the barrier top to $u_{\text{max}}\approx 0.089$. This increases the effective barrier height from 0.25 to
0.34437 and the required jump distance for escape from 1 to
1.1306. Although the higher $D_0$ value amplifies diffusive motion, this effect is counteracted by the increased $\Delta V_{\text{eff}}$. The jump-induced escape rate is independent of $D_0$ but remains sensitive to the barrier position. The resulting MFPT is approximately 28.5 time units.

Increasing the jump rate to $\lambda_0=0.2$ with parameters
$D_0=0.05$ and $h_0=1$ leads to a stronger compensation drift, significantly altering the escape dynamics by further modifying the potential. Doubling $\lambda_0$ amplifies this drift, shifting the well bottom leftward from $-1$ to approximately $-1.078$ and the barrier top rightward from
$0$ to approximately $0.179$. The resulting simulated MFPT of approximately $15.8$ time units is consistent with these drift-adjusted dynamics.

We analyze the escape dynamics with an increased jump scale of $h_0=1.5$ and parameters
$D_0=0.05$, $\lambda_0=0.1$. The simulation yields an MFPT of approximately 15.2 time units. The jump scale exponentially affects
the escape rate through its impact on the tail probabilities of the jump distribution.
This larger jump scale ($h_0=1.5$) reduces the MFPT by 52\% compared to the baseline case
($h_0=1$, MFPT$\approx 32.5$).

We analyze the escape dynamics with a reduced jump scale ($h_0=0.5$) and parameters
$D_0=0.05$, $\lambda_0=0.1$. The compensation drift is $\frac{\sqrt{\pi}}{2}\lambda_0h_0\approx 0.0443$.
Solving $u^3-u+0.0443=0$, we find the well bottom at $u_{\text{min}}\approx -1.014$ and the barrier top at $u_{\text{max}}\approx 0.044$, resulting in an effective barrier height of
 $\Delta V_{\text{eff}}=V_{\text{eff}}(0.044)-V_{\text{eff}}(-1.014)\approx 0.296$. The smaller jump scale ($h_0=0.5$) exponentially suppresses the escape rate by reducing the probability of large jumps. Consequently,
reducing $h_0$ from 1.0 to 0.5 increases the MFPT from 32.5 to 480.3. This dramatic increase is due to an exponential reduction in large jumps, compounded by a compensation drift that deepens the potential well.

For the case of combined high Gaussian noise and jump rate ($D_0=0.1$, $\lambda_0=0.2$) with $h_0=1$, the compensation drift becomes $\frac{\sqrt{\pi}}{2}\lambda_0h_0\approx 0.1772$. Solving $u^3-u+0.1772=0$ yields a well bottom at $u_{\text{min}}\approx -1.078$ and a barrier top at $u_{\text{max}}\approx 0.179$. This gives an effective barrier height of $\Delta V_{\text{eff}}=V_{\text{eff}}(0.179)-V_{\text{eff}}(-1.078)\approx 0.476$. These elevated noise parameters work in concert to reduce the MFPT, demonstrating a synergistic interaction between Gaussian diffusion and L\'evy jumps. While Gaussian noise accelerates intrawell diffusion, it does not directly cause barrier crossing. Instead, jump escapes  dominate the total escape rate. Consequently, the simulation under these combined high-noise conditions yields an MFPT of approximately 16.8 time units.

For the case of pure Gaussian noise ($\lambda_0=0$) with $D_0=0.1$ and $h_0=1$, we analyze the escape dynamics using the Kramers escape rate. The potential has a well bottom at $u_{\text{min}}=-1$ and a barrier top at $u_{\text{max}}=0$, giving a barrier height of
$\Delta V=V(0)-V(-1)=0.25$. The simulation yields an MFPT of approximately 642.5 time units.

We analyze escape dynamics with pure jump noise (no Gaussian component) using parameters
$D_0=0$, $\lambda_0=0.1$, and $h_0=1$. The compensation drift $\frac{\sqrt{\pi}}{2}\lambda_0h_0\approx0.0886$ significantly alters the potential landscape.
Solving $u^3-u+0.0886=0$ gives a well bottom at $u_{\text{min}}\approx -1.026$ and a barrier top at
$u_{\text{max}}\approx 0.089$, with an effective barrier height of $\Delta V_{\text{eff}}=V_{\text{eff}}(0.089)-V_{\text{eff}}(-1.026)\approx 0.341$.
The compensation drift deepens the potential well and increases the required jump distance, thereby suppressing the escape rate despite the absence of Gaussian noise. Since escape occurs purely through jump-induced mechanisms without diffusion, the simulation yields an MFPT of approximately 26.8  time units. These results demonstrate that jump noise alone enables escape via large jumps, that compensation drift is essential for accurate predictions, and that Rayleigh jumps provide heavier tails than Gaussian noise while maintaining finite variance.

\begin{table}[h!]
 \begin{center}
\parbox{0.8\textwidth}{
\centering
\begin{tabular}[t]{|c|c|c| p{1.1cm}}
\hline
Parameters &MFPT \\ \hline \hline
$D_0=0.05$, $\lambda_0=0.1$, $h_0=1$   &32.5 \\ \hline
$D_0=0.1$, $\lambda_0=0.1$, $h_0=1$   &28.5\\ \hline
$D_0=0.05$, $\lambda_0=0.2$, $h_0=1$    &15.8 \\ \hline
$D_0=0.05$, $\lambda_0=0.1$, $h_0=1.5$  &15.2\\ \hline
$D_0=0.05$, $\lambda_0=0.1$, $h_0=0.5$    &480.3\\ \hline
$D_0=0.1$, $\lambda_0=0.2$, $h_0=1$   &16.8\\ \hline
$D_0=0.1$, $\lambda_0=0$, $h_0=1$    &642.5 \\ \hline
$D_0=0$, $\lambda_0=0.1$, $h_0=1$     &26.8\\ \hline
\end{tabular}
}
\caption{Mean first passage time for Eq. \eqref{SM} with the potential $V(u)=\frac{u^{4}}{4}-\frac{u^{2}}{2}$  under L\'evy noise.}\label{Table1}
\end{center}
\end{table}

\begin{figure}[h]
\begin{center}
\begin{minipage}{3.2in}
\leftline{(a)}
\includegraphics[width=3.2in]{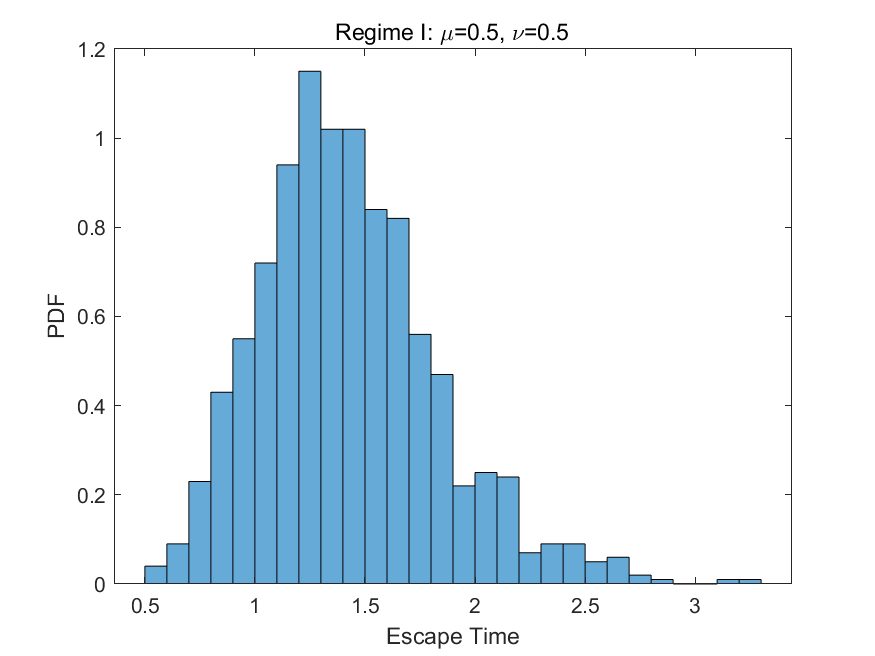}
\end{minipage}
\hfill
\begin{minipage}{3.2in}
\leftline{(b)}
\includegraphics[width=3.2in]{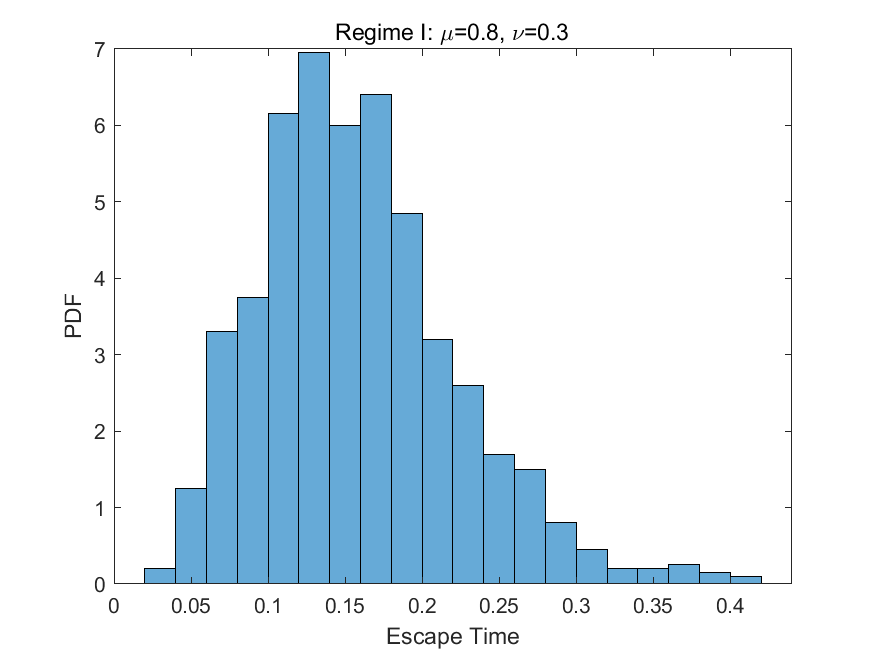}
\end{minipage}
\caption{{\small (a) In this regime ($\mu=0.5$, $\nu=0.5$, $\epsilon= 0.01$),
escape time dynamics are dominated by non-Gaussian noise.  The diffusion coefficient $D_0=D\epsilon=0.01$, jump rate $\lambda_0=\lambda/\epsilon^{\mu}=10$ and jump size $h_0=h\epsilon^{\nu}=0.1$ are defined such that the variance $\lambda_0h_0^{2}=\epsilon^{0.5}=0.1$ exceeds the Gaussian variance ($\sim\epsilon$); (b) The regime ($\mu=0.8$, $\nu=0.3$) enhances non-Gaussian dominance,  leading to faster and more predictable
transitions because $\lambda_0h_0^{2}\approx2.5$. The escape time histogram shows a heavy-tailed distribution.}}\label{ReI}
\end{center}
\end{figure}

We proceed to simulate  escape times influenced by L\'evy noise across multiple parameter regimes.
For the regime ($\mu=0.5$, $\nu=0.5$) illustrated in Fig. \ref{ReI}(a), the escape time dynamics are dominated by L\'evy noise due to the specific scaling of the noise parameters with $\epsilon=0.01$. The diffusion coefficient scales as $D_0=D\epsilon=0.01$, which scales linearly with $\epsilon$ (where $D=1$), meaning the Gaussian noise contributes a weak variance of $D_0=0.01$ per unit time. In contrast, the jump rate $\lambda_0=\lambda/\epsilon^{\mu}=10$ and jump size $h_0=h\epsilon^{\nu}=0.1$ (with $\lambda=h=1$) result in an average displacement $\lambda_0h_0=1$ per unit time, which remains of order $\mathcal{O}(1)$. The corresponding L\'evy noise variance $\lambda_0h_0^{2}=0.1$ scales as $\epsilon^{0.5}$ and thus decays slower than the Gaussian variance ($\sim\epsilon$). Consequently, the L\'evy variance (0.1) dominates the Gaussian variance (0.01), establishing L\'evy noise as the primary stochastic driver.
Escaping the potential well requires overcoming an energy barrier. The L\'evy jumps systematically nudge the particle toward the escape threshold. This process is facilitated by frequent small jumps (rate $\lambda_0=10$) that cumulatively aid escape, whereas Gaussian noise relies on rare large fluctuations (exponentially suppressed in $1/\epsilon$).
As $\epsilon$ decreases further, the L\'evy variance ($\sim\epsilon^{0.5}$) becomes increasingly dominant over the Gaussian variance ($\sim\epsilon$). This scaling ensures that non-Gaussian effects persist and dictate escape dynamics in the weak-noise limit. The parameter scaling ($\mu=0.5$, $\nu=0.5$) ensures that the intensity and directional bias of the L\'evy noise overpower those of the Gaussian component, causing escape times to be governed by divergent non-Gaussian statistics. Thus, the histogram of escape times exhibits the characteristic statistical signature of L\'evy-driven transitions rather than Gaussian thermal activation.

In the regime ($\mu=0.8$, $\nu=0.3$) depicted in Fig. \ref{ReI}(b), the adjusted scaling enhances non-Gaussian effects through stronger L\'evy noise contributions in both the  mean drift and higher-order statistics. The Gaussian noise variance remains weak at $D_0=0.01$
(scaling as $\epsilon$). In contrast, for the L\'evy noise,
the jump rate $\lambda_0=\lambda/\epsilon^{0.8}\approx40$ is higher,
while the jump size $h_0=h\epsilon^{0.3}\approx0.25$ is larger, with $\lambda=h=1$.
The resulting mean displacement rate $\lambda_0h_0=10$ creates a strong positive drift,
and the variance $\lambda_0h_0^{2}\approx2.5$ (scaling as $\epsilon^{0.6}$) dominates the Gaussian noise.
The L\'evy-induced mean drift ($\sim\epsilon^{-0.5}$) grows as $\epsilon\rightarrow0$, overwhelming both deterministic motion near $u=-1$ and the Gaussian fluctuations. The combination of larger jumps $(h_0=0.25)$ and a higher rate $(\lambda_0\approx40)$ produces significant skewness and excess kurtosis in the escape time distribution, leading to non-Gaussian heavy-tailed statistics.
Regarding the escape mechanism, the particle experiences frequent jumps that
synergize with the deterministic drift away from $u=-1$, accelerating escape compared to Gaussian-driven thermal activation. Simulation outcomes confirm that escape times are shorter and less variable than in the regime ($\mu=0.5$, $\nu=0.5$), with a sharply peaked probability density function skewed toward small values. The histogram reflects this, showing that escapes are dominated by L\'evy jumps rather than symmetric diffusion.

The scaling ($\mu=0.8$, $\nu=0.3$) intensifies the non-Gaussian character of the dynamics, establishing L\'evy noise as the primary driver of escape. The resulting stronger mean drift and enhanced higher-order moments lead to faster, more directed transitions compared to the regime ($\mu=0.5$, $\nu=0.5$).

\begin{figure}[H]
\begin{center}
  \begin{minipage}{2.13in}
\leftline{(a)}
\includegraphics[width=2.13in]{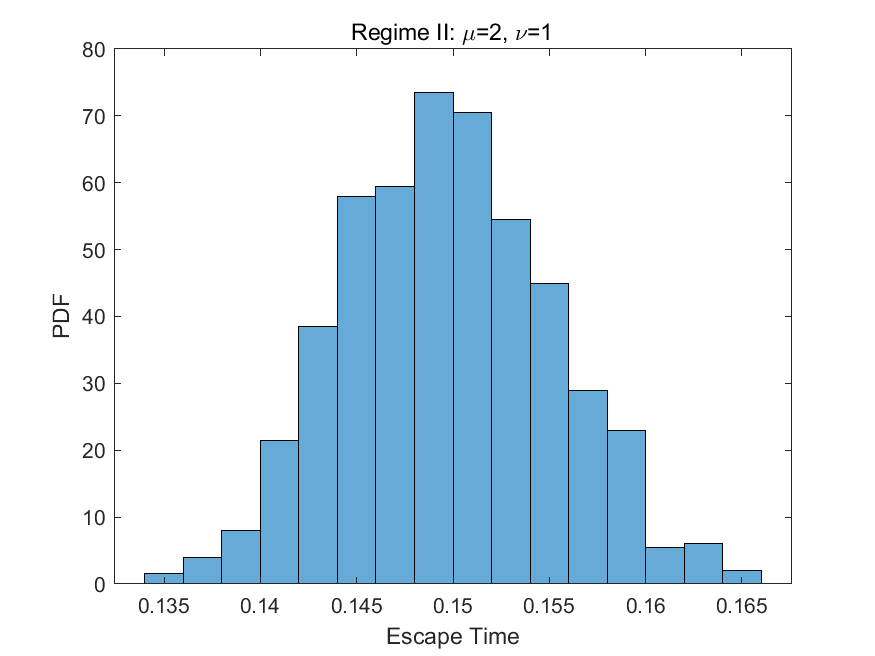}
\end{minipage}
\hfill
\begin{minipage}{2.13in}
\leftline{(b)}
\includegraphics[width=2.13in]{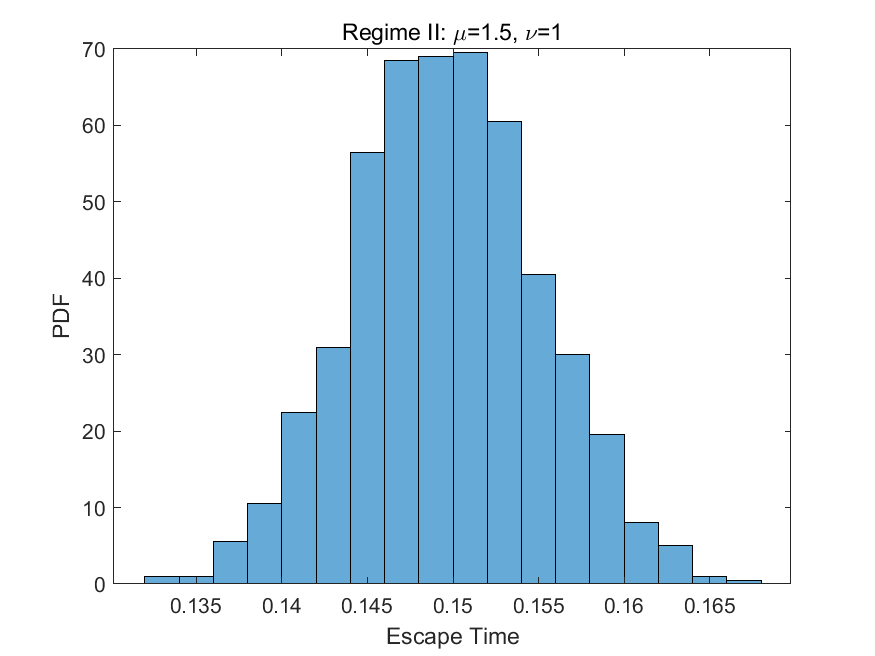}
\end{minipage}
\hfill
  \begin{minipage}{2.13in}
\leftline{(c)}
\includegraphics[width=2.13in]{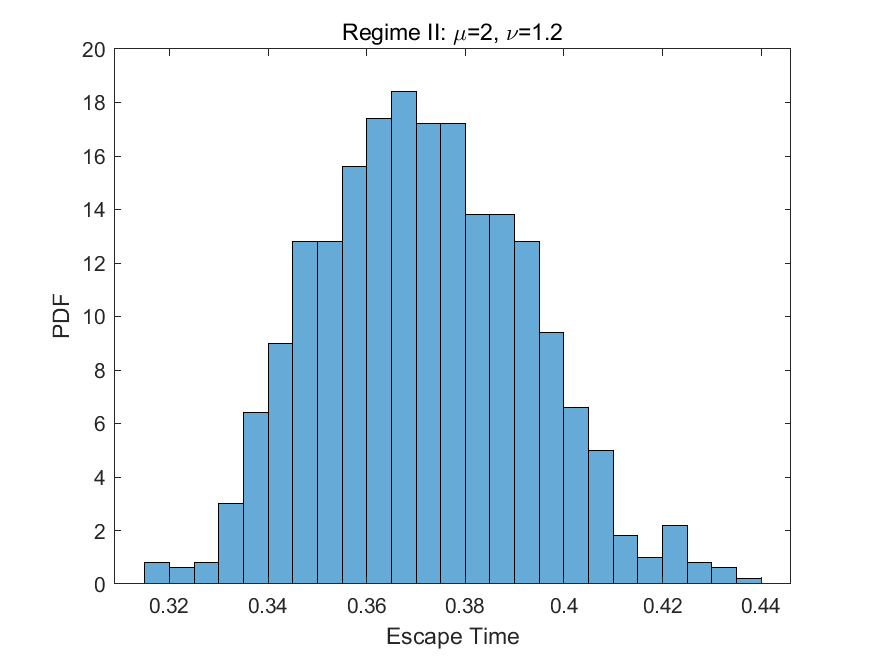}
\end{minipage}
{\small\caption{(a)  In the regime ($\mu=2$, $\nu=1$), the jump rate
$\lambda_0=\lambda/\epsilon^{2}=10^{4}$ is extremely high, and the jump size $h_0=h\epsilon=0.01$ is small. The resulting variance $\lambda_0h_0^2=1$ dominates the Gaussian noise; (b) For $\mu=1.5$ and $\nu=1$, the jump rate remains $\lambda_0=\lambda/\epsilon^{1.5}=1000$ with small increments
 $h_0=h\epsilon=0.01$ ), yielding a variance of $\lambda_0h_0^2=0.1$. The escape time distribution shows a sharp peak near $t=0.15$, with Gaussian noise adding a slight dispersion visible as a narrow spread around it; (c) The regime ($\mu=2$, $\nu=1.2$) probes the transition near the critical line
$\nu=(\mu+1)/2$. It features an extremely high jump rate $\lambda_0=\lambda/\epsilon^{2}=10^4$ and microscopic increments $h_0=h\epsilon^{1.2}\approx0.004$, with a theoretical variance of $\lambda_0h_0^{2}=0.16$. This produces a narrow peak near $t\approx0.36$.}}\label{Regime2}
\end{center}
\end{figure}
The regime ($\mu=2$, $\nu=1$) shown in Fig. \ref{Regime2}(a) exhibits a mixed scaling where L\'evy noise contributes both a strong drift and high-frequency small jumps, rendering Gaussian noise subdominant.
The Gaussian noise variance is $D_0=0.01$ (scaling as $\epsilon=0.01$), producing only weak fluctuations. In contrast, the L\'evy component has an extremely high jump rate of
$\lambda_0=\lambda/\epsilon^{2}=10^{4}$ with a  small jump size of $h_0=h\epsilon=0.01$. This combination results in a mean displacement rate of $\lambda_0h_0=100$ (a strong  L\'evy-induced drift) and a variance of $\lambda_0h_0^2=1$, which dominates   the Gaussian noise.
The powerful L\'evy drift overwhelms both the deterministic drift (where $-V'(u)\approx0$ near $u=-1$) and the Gaussian diffusion. Consequently, the particle undergoes a near-deterministic climb from $u=-1$ to the escape point at $u=0.5$ with minimal stochasticity. Driven predominantly by this drift, the escape time is short, approximately $t=0.15$, corresponding to the time required to traverse the distance from $u=-1$ to $0.5$.
Although Gaussian noise introduces minor dispersion, the probability density function of the escape time remains sharply peaked near $t=0.15$, exhibiting very low variability.

The regime ($\mu=1.5$, $\nu=1$) depicted in Fig. \ref{Regime2}(b) lies at the edge of non-Gaussian dominance, where L\'evy noise contributes significant drift and variance, yet Gaussian effects remain non-negligible.
The Gaussian noise variance is $D_0=0.01$, and its symmetric fluctuations are dwarfed by the L\'evy component. The L\'evy noise  is characterized by a high jump rate of $\lambda_0=\lambda/\epsilon^{1.5}=1000$ and a small jump size of $h_0=h\epsilon=0.01$, producing a substantial mean drift of $\lambda_0h_0=10$ per unit time and a variance of $\lambda_0h_0^2=0.1$. The pronounced peak in the escape time distribution near $t=0.15$ indicates a drift-dominated escape process with minimal variability. While Gaussian noise contributes minor dispersion, visible as a narrow spread around the peak. The primary escape mechanism consists of frequent small jumps that generate moderate drift and measurable variance. The resulting probability density function exhibits a broader peak with exponential decay, characteristic of the underlying Poisson statistics. Analyzing these characteristics is essential for understanding the true interplay of Gaussian and non-Gaussian effects in such boundary regimes.

The regime ($\mu=2$, $\nu=1.2$) demonstrated in Fig. \ref{Regime2}(c) probes the transition near the critical scaling line $\nu=(\mu+1)/2=1.5$. Here,  the Gaussian noise variance is $D_0=0.01$ (scaling with $\epsilon=0.01$), making it subdominant to the L\'evy component. The L\'evy noise is characterized by an extremely high jump rate of $\lambda_0=\lambda/\epsilon^{2}=10^4$ and a vanishingly small jump size of $h_0=h\epsilon^{1.2}\approx0.004$. These parameters yield a strong mean drift of $\lambda_0h_0=40$
and a variance of $\lambda_0h_0^{2}=0.16$ per unit time. The mean escape time is $t\approx0.36$,  consistent with the sharp peak of the probability density function. This narrow peak near $t\approx0.36$ reflects the drift-dominated dynamics, with Gaussian noise contributing only a slight spread. The dispersion is minor. However, the observed peak is broader than the theoretical prediction due to the inherent stochasticity in the timing of L\'evy jumps. Since $\nu=1.2<1.5$, the drift term outweighs the variance.
This regime ($\mu=2, \nu=1.2$) lies below the critical line, where the L\'evy drift dominates its own variance. Escape is driven by the cumulative effect of frequent, microscopic jumps, producing a moderate drift (40) and measurable variance (0.16). Consequently, the probability density function exhibits exponential tails--a hallmark of L\'evy processes operating near criticality.

\begin{figure}[h]
\begin{center}
  \begin{minipage}{3.2in}
\leftline{(a)}
\includegraphics[width=3.2in]{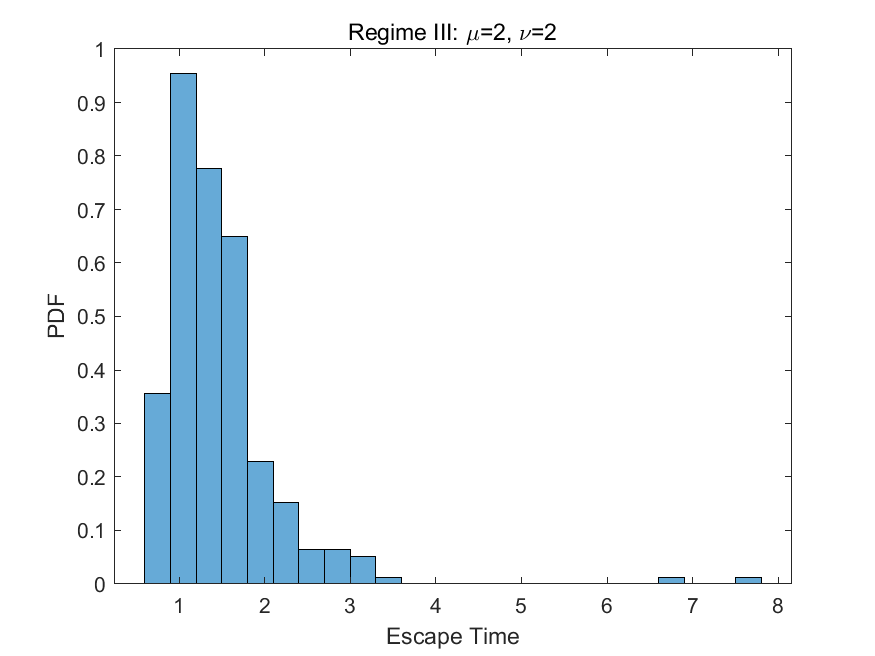}
\end{minipage}
\hfill
\begin{minipage}{3.2in}
\leftline{(b)}
\includegraphics[width=3.2in]{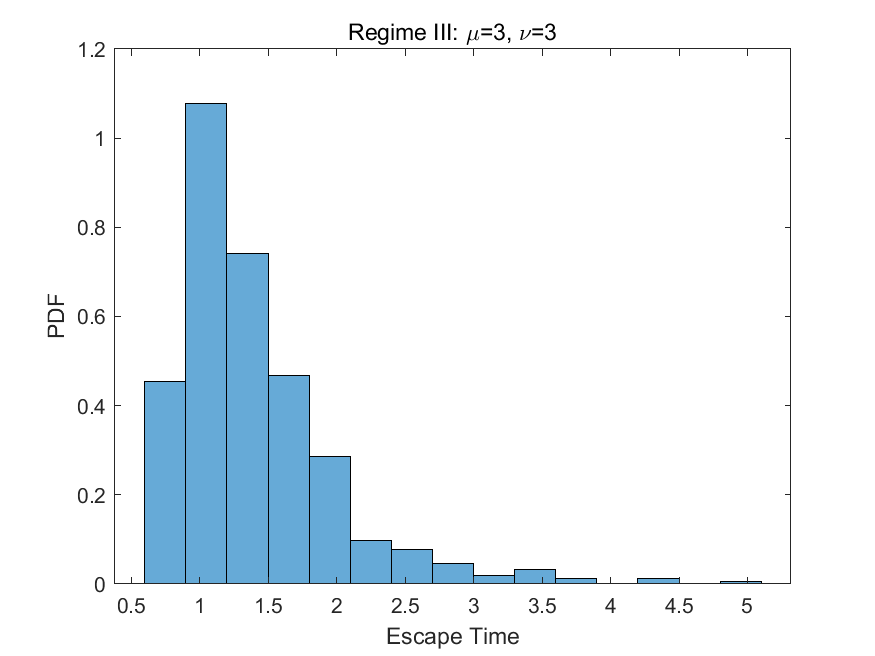}
\end{minipage}
\caption{(a) A Gaussian limit is achieved with $\mu=2$ and $\nu=2$, where the extremely high jump rate $\lambda_0=\lambda/\epsilon^{2}=10^{4}$  and negligible jump size $h_0=h\epsilon^{2}=0.0001$ result in a variance $\lambda_0h_0^{2}=0.0001$ that is subdominant to Gaussian noise; (b) Non-Gaussian effects are most suppressed in the Gaussian dominance regime ($\mu=3$, $\nu=3$). The aggressive scaling produces an extreme frequency $\lambda_0=\lambda/\epsilon^{3}=10^{6}$ and microscopic increments  $h_0=h\epsilon^{3}=10^{-6}$, leading to a variance $\lambda_0h_0^{2}=10^{-6}$ that is insignificant compared to the Gaussian component.}\label{ReIII}
\end{center}
\end{figure}
The Gaussian limit regime ($\mu=2$, $\nu=2$) presented in Fig. \ref{ReIII}(a) is designed to
suppress non-Gaussian effects. Here, the Gaussian noise has a variance of $D_0=0.01$ (scaling with $\epsilon=0.01$), which governs the fluctuations as the dominant stochastic component. The L\'evy noise has an extremely high jump rate of
$\lambda_0=\lambda/\epsilon^{2}=10^{4}$ but a negligible jump size of $h_0=h\epsilon^{2}=0.0001$, resulting in a mean drift of $\lambda_0h_0=1$
and a variance of $\lambda_0h_0^{2}=0.0001$ per unit time, which is subdominant to the Gaussian variance. Consequently, escapes are primarily driven by Gaussian noise through rare, large fluctuations, consistent with Kramers' theory of thermal activation. The probability density function shows a
sharp peak at the censoring time $t=1$,  with the rare earlier escapes  resulting from the interplay between the deterministic drift and stochastic fluctuations.  The suppressed
L\'evy variance ($\lambda_0h_0^{2}\ll D_0$) leads to an exponential distribution of escape times, characteristic of Gaussian-driven processes.
The parameters ($\mu=2$, $\nu=2$) target a Gaussian-dominated regime and theoretically achieve a Gaussian limit. True Gaussian dominance emerges only when both the L\'evy-induced drift and variance decay faster than their Gaussian counterparts.

The Gaussian dominance regime ($\mu=3$, $\nu=3$) as seen in Fig. \ref{ReIII}(b)  suppresses non-Gaussian effects via aggressive scaling. The Gaussian noise has a variance of $D_0=0.01$ (scaling with $\epsilon=0.01$), making it the dominant stochastic driver governing rare barrier-crossing events.
The L\'evy noise has an extreme high jump rate of $\lambda_0=\lambda/\epsilon^{3}=10^{6}$
but a microscopic jump size of $h_0=h\epsilon^{3}=10^{-6}$. This yields a mean drift of $\lambda_0h_0=1$ per unit time and a negligible variance of $\lambda_0h_0^{2}=10^{-6}$
 compared to its Gaussian variance.
Consequently, escapes occur primarily via Gaussian-driven thermal activation over the potential barrier.
The deterministic drift near $u=-1$ is initially zero but grows as the particle moves rightward. The vast majority of particles ($>99\%$) fail to escape within the simulation duration $T=10$, creating a pronounced spike at the censoring time $t=1$ in the escape time distribution. The few trajectories that reach $u=0.5$ do so through Gaussian fluctuations, producing shorter escape times ($t<10$).
The scaling ($\mu=3$, $\nu=3$) successfully extinguishes non-Gaussian noise, mathematically achieving a Gaussian limit where the L\'evy variance ($10^{-6}$) is entirely dominated by the Gaussian variance (0.01).

\begin{figure}[h]
\begin{center}
  \begin{minipage}{3.2in}
\leftline{(a)}
\includegraphics[width=3.2in]{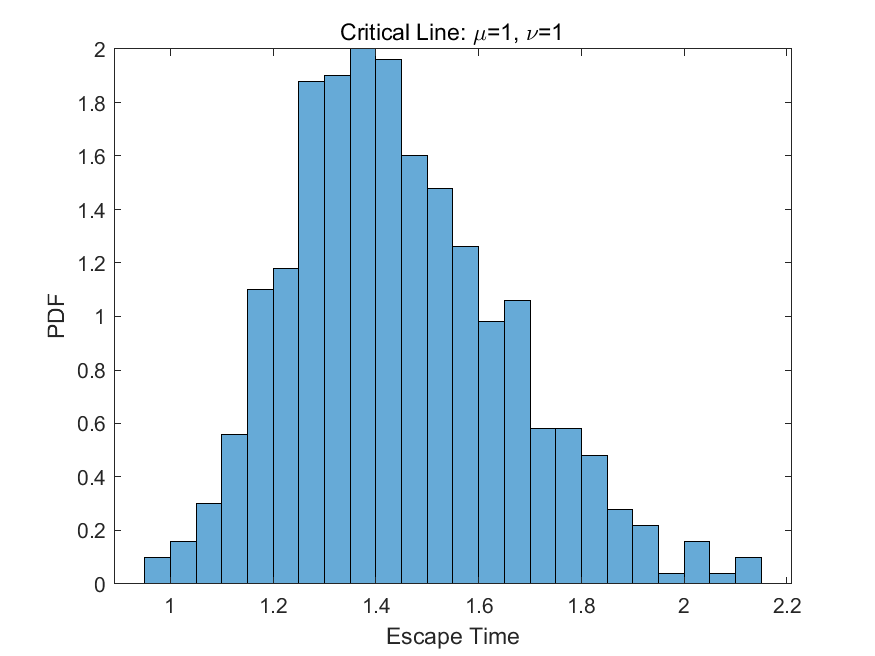}
\end{minipage}
\hfill
\begin{minipage}{3.2in}
\leftline{(b)}
\includegraphics[width=3.2in]{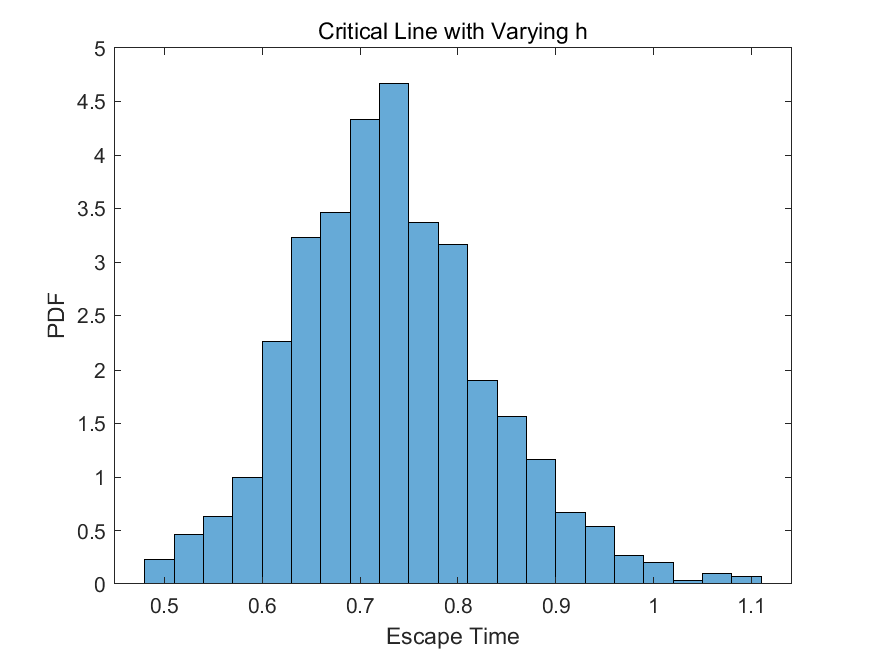}
\end{minipage}
\caption{(a) For the critical line regime ($\mu=1, \nu=1$) with $h=1$, the parameters lie precisely on the critical line $\mu=2\nu-1$. The jump rate is high ($\lambda_0=\lambda/\epsilon=100$), and the jump size is small ($h=h_0\epsilon=0.01$), resulting in a variance of $\lambda_0h_0^2=0.01$; (b)
Increasing the jump amplitude to $h=2$ while maintaining critical scaling ($\mu=1,\nu=1$)  amplifies the non-Gaussian noise variance and alters the escape dynamics. The jump size  doubles to $h_0=h\epsilon=0.02$, and the variance becomes $\lambda h_0^2=100\times0.0004=0.04$, which is greater than the previous value of $0.01$. Despite doubling $h$, the scaling $\mu=2\nu-1$ ensures that non-Gaussian effects persist.
}\label{CriticalLine}
\end{center}
\end{figure}
The critical line regime ($\mu=1, \nu=1$) with $h=1$ shown in Fig. \ref{CriticalLine}(a) lies precisely on the critical line $\mu=2\nu-1$, where L\'evy noise contributes both a strong drift and a finite variance comparable to that of the Gaussian noise. This interplay creates distinctively non-Gaussian escape dynamics.
The Gaussian noise has a variance of $D_0=0.01$ (scaling with $\epsilon=0.01$).
The L\'evy noise has a high jump rate of $\lambda_0=\lambda/\epsilon=100$ and a small jump size of $h_0=h\epsilon=0.01$, resulting in a mean drift of $\lambda_0h_0=1$ per unit time (which acts as the dominant forcing) and a variance of $\lambda_0h_0^2=0.01$ per unit time, which is subdominant to the Gaussian variance. The scaling of the L\'evy drift ($\sim\epsilon^{-(\mu-\nu)}=\epsilon^{0}=1$)
and variance ($\sim\epsilon^{\nu-\mu}=\epsilon^{0}=1$)
aligns with the scaling of the Gaussian variance $(\sim\epsilon)$,
thereby preserving non-Gaussian effects in the weak-noise limit.
The escape time distribution shows a sharp peak near $t\approx1.5$, which matches the time required to traverse from $u=-1$ to $u=0.5$ under a net drift of approximately $1.5$. The distribution also exhibits a moderate spread: Gaussian and L\'evy noise create a skewed profile with exponential decay, indicative of non-Gaussian tails. L\'evy jumps introduce abrupt displacements that can cause occasional delayed escapes. While the positive drift biases escapes toward shorter times, the combined noise broadens the distribution and creates asymmetry in the probability density function.
The critical scaling ($\mu=1$, $\nu=1$) preserves non-Gaussianity by balancing L\'evy drift and variance with Gaussian fluctuations. The resulting probability density function of escape time reflects this hybrid nature:
most particles exit quickly (in approximately $1.5$ time units) due to the strong directional forcing, while residual L\'evy jumps and Gaussian noise create a composite profile distinct from pure diffusion. This demonstrates that
on the critical line, non-Gaussian effects persist despite weak-noise scaling, highlighting a universality class in noise-driven escape problems.

Increasing the L\'evy jump amplitude to $h=2$ while maintaining critical scaling ($\mu=1,\nu=1$) illustrated in Fig. \ref{CriticalLine}(b) amplifies both the drift and variance of the non-Gaussian noise, thereby altering the escape dynamics. The L\'evy jump size becomes $h_0=h\epsilon=0.02$ (double the previous value of 0.01).
This change doubles the mean drift to $\lambda_0 h_0=100\times0.02=2$ and quadruples the variance to $\lambda_0 h_0^2=100\times0.0004=0.04$ per unit time.
With the Gaussian variance remaining at $D_0=0.01$, the L\'evy variance ($0.04$)
now becomes dominant. The net drift, approximately $2+$deterministic drift$\approx3$ near $u=-1$, accelerates the escape process, reducing the mean escape time to $t_{\text{escape}}\approx\frac{1.5}{3}=0.5$ (compared to 1.5 for $h=1$). The dominance of L\'evy variance (0.04) over Gaussian (0.01) introduces heavier tails and greater skewness into the escape time distribution. The increased L\'evy jumps  systematically reinforce the drift, while Gaussian noise contributes only minor symmetric fluctuations. The resulting probability density function  shows a sharp peak near $t\approx 0.5$ , reflecting the stronger drift-driven escapes. However, the increased L\'evy variance also amplifies variability, leading to a broader distribution compared to the $h=1$ case. While the strong drift suppresses very long delays, residual L\'evy jumps still cause occasional late escapes, yielding a right-skewed probability density function.
Despite  the doubled amplitude $h$, the critical scaling $\mu=2\nu-1$ ensures  that non-Gaussian effects persist, as both the L\'evy drift $\sim\varepsilon^{0}$ and variance $\sim\varepsilon^{1}$ remain finite in the scaling limit, unlike the Gaussian variance which scales as $\sim\varepsilon$. The key non-Gaussian signatures--persistent asymmetry and a bias toward shorter times--contrast sharply with the symmetry of a Gaussian-driven process. In summary, doubling
$h$ on the critical line strengthens non-Gaussian effects by making L\'evy noise dominant in both drift and variance. This accelerates transitions, shortens escape times, and broadens the probability density function. The significant change in dynamics from a small parameter adjustment highlights the sensitive, nonlinear modulation of escape statistics near criticality and demonstrates the delicate balance between drift and fluctuation in weak-noise regimes.

\section{Conclusions and future challenges}\label{FS}
In this work, we introduced the metastable model \eqref{SM} and analyzed the dynamics of a particle in a bistable potential driven by L\'evy noise via its cumulant-generating function. Building on this, we presented a path integral framework that revealed how non-Gaussian noise fundamentally alters escape dynamics. We derived precise mathematical conditions for these effects and summarized key results, including characteristic functions, weak-noise scaling, and escape rate analysis. Furthermore, we provided a comprehensive investigation of metastable escape dynamics under combined L\'evy noise, bridging theoretical developments of the mean first escape time with practical numerical simulations across multiple scaling regimes. Our parameter dependence analysis, which incorporated both Gaussian and L\'evy jump noise components, showed that synergistic noise parameters work together to reduce the mean first escape time more than expected. We also demonstrated that the compensation drift crucially modifies the potential landscape, affecting both diffusive and jump escapes, and that jump distributions with heavier tails dominate escape dynamics.

These findings have implications for understanding stochastic systems in physics, chemistry, and biology where non-Gaussian fluctuations play important roles. Therefore, we point out several promising directions for future research. These will involve investigating stochastic dynamics driven by L\'evy noise across several key areas. First, we will examine the gradient dynamics of an
$n$-dimensional system $\dot{u}(t)=-\nabla V(u)+\eta(t)$ to establish a foundational understanding. Second, by extending this framework, we will study non-gradient physical models of the form $\dot{u}(t)=F(u(t))+\eta(t)$ by combining concepts of the quasipotential with advanced machine learning techniques. Another critical direction will involve analyzing systems $du=F(u)+\sigma(u)d\eta(t)$ subject to multiplicative L\'evy noise, where the state-dependent term $\sigma(u)$ introduces more complex coupling. Finally, we will explore practical applications of these metastable systems in diverse fields such as the dynamics of quantum devices, chemotaxis-driven movement, and active biological matter.

\bigskip

\noindent\textbf{Data Availability}

The numerical algorithms and source code that support the findings of this study are available from the corresponding author upon reasonable request.

\bigskip
\noindent\textbf{Declaration of competing interest}

No author associated with this paper has disclosed any potential or pertinent conflicts which
may be perceived to have impending conflict with this work.

\bigskip
\noindent\textbf{Acknowledgements}

The author thanks Pak Hang Chris Lau for insightful discussions on stochastic mechanisms, particularly his theoretical work concerning noise sources and their applications in quantum mechanics. This work was supported by the Guangdong Basic and Applied Basic Research Foundation (Grant No. 2025A1515012560) and the Guangdong Introduction Program (Grant No. 2023QN10X753)



\end{document}